# A Concept Annotation System for Clinical Records

Ning Kang, Rogier Barendse, Zubair Afzal, Bharat Singh, Martijn J. Schuemie,
Erik M. van Mulligen, and Jan A. Kors

Dept. of Medical Informatics, P.O. Box 2040
Erasmus University Medical Center, 3000 CA   Rotterdam, The Netherlands
{n.kang, r.barendse, z.afzal, b.singh, m.schuemie, e.vanmulligen, j.kors}@erasmusmc.nl

**Abstract.** Unstructured information comprises a valuable source of data in clinical records. For text mining in clinical records, concept extraction is the first step in finding assertions and relationships. This study presents a system developed for the annotation of medical concepts, including medical problems, tests, and treatments, mentioned in clinical records. The system combines six publicly available named entity recognition system into one framework, and uses a simple voting scheme that allows to tune precision and recall of the system to specific needs. The system provides both a web service interface and a UIMA interface which can be easily used by other systems. The system was tested in the fourth i2b2 challenge and achieved an F-score of 82.1% for the concept exact match task, a score which is among the top-ranking systems. To our knowledge, this is the first publicly available clinical record concept annotation system.

## 1     Introduction

Clinical concepts such as medical problems, tests, and treatments, are important types of clinical data in electronic medical record systems. Obtaining accurate clinical concepts is a common and critical task for clinical research and care, and a necessary initial step in finding clinical assertions and relationships. In this paper, we describe a concept annotation system for clinical records. The system was tested on the concept extraction task in the fourth i2b2 challenge on clinical records (1). By providing both a web service and a UIMA (Unstructured Information Management Architecture) interface (2), the system is easily integrated with other systems.

## 2     Methods

Our system for the extraction of problems, tests, and treatments from plain clinical record text consists of tagging the data with a variety of named entity recognizers and chunkers, and combining the resulting annotations of these systems into a final annotation set by a simple voting scheme.

## 2.1 Definition of concept types

In our annotation system, problems are phrases that contain observations made by patients or clinicians about the patient's body or mind that are thought to be abnormal or caused by a disease. Treatments are phrases that describe procedures, interventions, and substances given to a patient in an effort to resolve a medical problem. Tests are phrases that describe procedures, panels, and measures that are done to a patient or a body fluid or sample in order to discover, rule out, or find more information on a medical problem (1). An example annotation is: "The patient had [increasing dyspnea]PROBLEM on exertion, he had [a bronchoalveolar lavage]TREATMENT performed, and [CBC]TEST was unremarkable."

## 2.2 Concept annotation systems

We selected six annotation systems of which the output was to be combined. One was a locally developed concept recognition and normalization tool, called Peregrine (3), which used the UMLS as a dictionary. The other five systems were publicly available named entity recognizers and chunkers, which could be downloaded directly from their official websites (ABNER 1.5, http://pages.cs.wisc.edu/~bsettles/abner; Lingpipe 3.8, http://alias-i.com/lingpipe; OpenNLP Chunker 2.1 and OpenNLPNer 2.1, http://opennlp.sourceforge.net; StanfordNer 1.1, http://nlp.stanford.edu/software/CRF-NER.shtml). The OpenNLP Chunker was used in combination with modules for sentence splitting, token annotation and part-of-speech tagging from the OpenNLP toolbox. All systems except Peregrine, were trained on the i2b2 training corpus, consisting of 349 clinical records with concept annotations.

## 2.3 Concept annotation steps

The following processing steps were done to generate the concept annotations (Figure 1):

1. All six tools were integrated into the UIMA framework.

2. The system provides a web service and a UIMA interface for other systems or clients to use; the input data are plain-text clinical records. When data are received from the clients, the system calls each of the six tools to annotate the records.

3. The annotation results of the individual systems are combined by a simple voting scheme. For each annotation, the number of systems that exactly match the annotation (i.e., the same start and end position, with the same concept type) is counted. If the count is larger or equal than a preset voting threshold, the annotation is considered to be annotated by the combined system, otherwise it is not annotated.

4. The combined annotations are output according to the i2b2 annotation file format, and sent back to clients via the web service or UIMA interface.

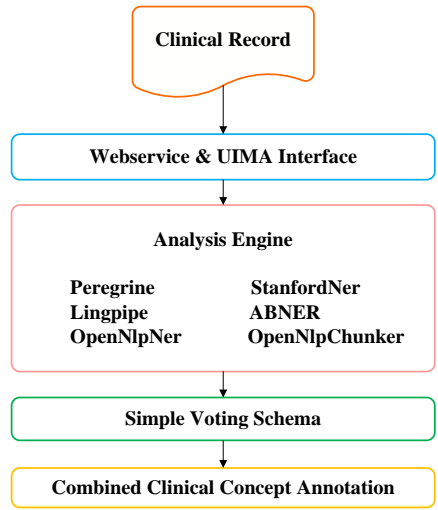

**Fig. 1.** Components of the concept annotation system.

## 3 Results

Our system participated in the fourth i2b2 challenge and achieved an F-score of 82.1% for the concept exact match task, which is among the top-ranking systems. For the errors, about 2% are caused by system annotations that have a wrong concept type, 8% have either a wrong start position or a wrong end position, and 8% have both a wrong start position and a wrong end position. The F-score for each of the three concept types was also calculated (data not shown). Differences between F-scores were at most 1.8%.

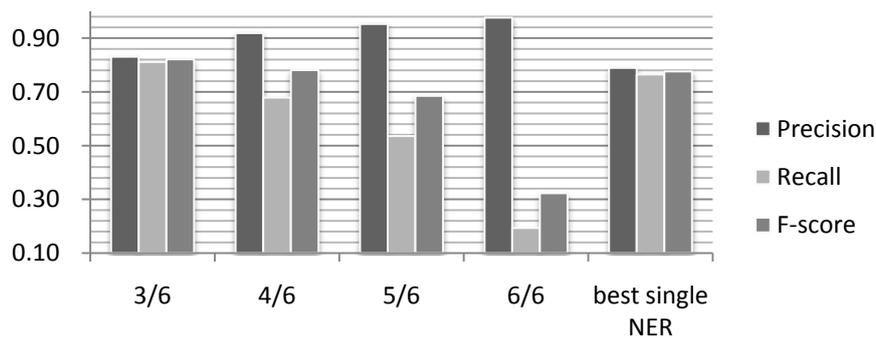

**Fig. 2.** Performance of the combined annotation on the i2b2 test corpus for varying voting thresholds between the six annotation systems.

Varying the voting threshold allows for different precision-recall settings (Figure 2). For a threshold of 3, precision and recall of the combined system is higher than that of the best individual system, OpenNLPNer, and the F-score increases by 4.5 percentage points. When we increase the threshold for agreement from 3 to 6, precision increases and recall drops.

## 4    Discussion

Our results for the i2b2 concept annotation task indicate that the combined annotation of a variety of annotation systems yields an F-score that is considerably higher than the best single system. Our system offers the possibility to vary precision and recall of the combined annotation by varying the voting threshold (cf. Figure 1). For example, a threshold of 5 would give a high precision (0.95) with a reasonable recall (0.53); with a threshold of 6 an even higher precision (0.98) would be possible, but at the expense of a poor recall (0.20).

## 5    Conclusion

To our knowledge, this is the first publicly available concept annotation system for clinical records. The combined annotation of clinical records annotated by different systems performs substantially better than any of the individual systems. The combination approach is straightforward and allows the balancing of precision versus recall. The UIMA interface of the system allows easy integration with other systems

**Acknowledgments.** This study was supported by the European Commission FP7 Program (FP7/2007-2013) under grant no. 231727 (the CALBC Project).